\documentclass[preprint,showpacs,amsmath,amsfonts,aps]{revtex4}

\begin{document}
\title{Extended symmetrical classical electrodynamics}

\author{A. V. Fedorov}
\email{anatoly.fedorov@gmail.com}
\author{E. G. Kalashnikov$^1$}
\address{$^1$Ulyanovsk State University, 42, Leo Tolstoy Str, Ulyanovsk, 432970,
 Russia}

\begin{abstract}
In the present article, we discuss a modification of classical
electrodynamics in which ``ordinary'' point charges  are absent. The modified
equations contain  additional terms describing the   induced charges and
currents.    The densities of the induced charges and currents   depend on the
vector $\mathbf k$ and the vectors of the electromagnetic field $\mathbf E$ and
$\mathbf B$.  It is shown that the vectors $\mathbf E$ and $\mathbf B$ can be
defined in terms of two 4-potentials and the components of $\mathbf k$ are the
components of the 4-tensor of the third rank. The Lagrangian 
 of modified electrodynamics is defined. The conditions are derived
at which only one 4-potential determines the behavior of the electromagnetic
field. It is also shown  that static modified electrodynamics can describe the
electromagnetic field in the inner region of the electric monopole.  In the
outer region of the electric monopole the electric field is governed by the
Maxwell equations. It follows from  boundary conditions at the interface between
the inner and outer regions of the monopole   that the vector $\mathbf k$ has a
discrete spectrum. The electric and magnetic fields, energy and angular momentum
of the monopole are found for different eigenvalues of $\mathbf k$.
\end{abstract}


\maketitle

\section{Introduction}

%opening
  In recent years, there has been a growing interest in the study of classical
Maxwell-Chern-Simons (MCS) electrodynamics. The fundamental equations of
MCS-electrodynamics are:
\begin{equation}
\nabla\times\mathbf E+\frac{1}{c}\frac{\partial\mathbf B}{\partial
t}=0,\label{1}
\end{equation}
\begin{equation}
\nabla\cdot\mathbf B=0,\label{2}
\end{equation}
\begin{equation}
\nabla\times\mathbf B-\frac{1}{c}\frac{\partial\mathbf E}{\partial
t}=\frac{4\pi}{c}\mathbf j_o-m\mathbf B-\mathbf k\times\mathbf E,\label{3}
\end{equation}
\begin{equation}
\nabla\cdot\mathbf E=4\pi\rho_o+\mathbf k\cdot\mathbf B,\label{4}
\end{equation}
where  $\rho_o$ and $\mathbf j_o$ are the ``ordinary'' charge and current
densities, respectively, and ``ordinary'' charges are considered as point
particles. The quantities $m$ and $\mathbf k$ have dimensions of inverse length
and ones can be considered either as the components of the 4-gradient of the
dynamic pseudoscalar (axion) field \cite{sikivie, huang,   wilczek,
gasperini,jacobs,carroll,masso} or as the components of a constant
4-pseudovector \cite{jackiw,carroll2,lehnert,lehnert2,hariton}. With $m=0$ the
set of the equations (\ref{1})-(\ref{4}) was obtained in  noncommutative
electrodynamics \cite{gamboa}.

The equations (\ref{3}) and (\ref{4}) can be written in the following form
\begin{equation}
\nabla\times\mathbf B-\frac{1}{c}\frac{\partial\mathbf E}{\partial
t}=\frac{4\pi}{c}\mathbf j_o+\frac{4\pi}{c}\mathbf j_i,\label{3a}
\end{equation}
\begin{equation}
\nabla\cdot\mathbf E=4\pi\rho_o+4\pi\rho_i,\label{4a}
\end{equation}
where
\begin{equation}
\mathbf j_i=-\frac{c}{4\pi}(m\mathbf B+\mathbf k\times\mathbf E),\label{ji} 
\end{equation}
\begin{equation}
\rho_i=\frac{1}{4\pi}\mathbf k\cdot\mathbf B.\label{qi} 
\end{equation}
As distinct from the Maxwell equations, the equations (\ref{3a})-(\ref{4a})
contain the additional quantities $\mathbf j_i$ and $\rho_i$ which can be
interpreted as follows \cite{sikivie}. The quantity $\mathbf j_i$ is the current
density induced by magnetic and electric fields and the quantity $\rho_i$ is the
charge density induced by the magnetic field.    From (\ref{ji}) and (\ref{qi})
it follows that the induced current  and  charge densities are the functions of
the vectors of the electromagnetic field. In such a manner, an electric charge
can be considered as the secondary property of the electromagnetic field. It is
necessary to note that the induced current is not connected with mechanical
motion of any point charges. The first term on the right-hand side of (\ref{ji})
has arisen in magnetohydrodynamics \cite{jackiw} and the second term has
appeared in the description of the Hall effect \cite{landau1}.

In the absence of ``ordinary'' sources, the Maxwell equations can be considered
as two sets of  equations when  one is transformed into the other if the
replacement $\mathbf E\rightarrow\mathbf B$, $\mathbf B\rightarrow-\mathbf E$ 
takes place. The equations of MCS-electrodynamics do not have  such symmetry.
Using the equations (\ref{1})-(\ref{4}) , we shall try to formulate  the
equations of electrodynamics which will have such symmetry.
At first we assume   that the equation for the curl of  $\mathbf B$ has the same
form as the equation (\ref{3}) but the  quantities $\rho_o$ and $\mathbf j_o$
are equal to zero. Thus the equation (\ref{3}) can be written in the form: 
\begin{equation}
\nabla\times\mathbf B-\frac{1}{c}\frac{\partial\mathbf E}{\partial t}=-m\mathbf
B-\mathbf k\times\mathbf E.\label{5}
\end{equation}
 Then we carry out the replacement $\mathbf E\rightarrow\mathbf B$, $\mathbf
B\rightarrow-\mathbf E$ in the equation (\ref{5}) and get the  equation for the
curl $\mathbf E$ :
\begin{equation}
\nabla\times\mathbf E+\frac{1}{c}\frac{\partial\mathbf B}{\partial t}=-m\mathbf
E+\mathbf k\times\mathbf B.\label{6}
\end{equation}

Suppose that the quantities $m$ and $\mathbf k$ are constant in time and space.
Computing the curls of the equations (\ref{5}) and (\ref{6}), we find the
equations:
\begin{equation}
\square\mathbf B+\frac{2}{c}\mathbf k\times\frac{\partial\mathbf B}{\partial
t}+(m^2+k^2)\mathbf B=\frac{2m}{c}\frac{\partial\mathbf E}{\partial
t}-2m\mathbf k\times\mathbf E+\nabla f_1+\mathbf kf_2,\label{7} 
\end{equation}
\begin{equation}
\square\mathbf E+\frac{2}{c}\mathbf k\times\frac{\partial\mathbf E}{\partial
t}+(m^2+k^2)\mathbf E=2m\mathbf k\times\mathbf B-
\frac{2m}{c}\frac{\partial\mathbf B}{\partial t}+\nabla f_2-\mathbf
kf_1,\label{8} 
\end{equation}
where $\square$ is the d'Alembertian:
\begin{equation}
\square=\nabla^2-\frac{1}{c^2}\frac{\partial^2}{\partial t^2}\label{9}
\end{equation}
and
\begin{equation}
f_1=\nabla\cdot\mathbf B-\mathbf k\cdot\mathbf E,\label{10}
\end{equation}
\begin{equation}
f_2=\nabla\cdot\mathbf E+\mathbf k\cdot\mathbf B.\label{11}
\end{equation}
Suppose that $f_1$ and $f_2$ are equal to zero. In this case the divergences
$\mathbf B$ and $\mathbf E$ can be written as:
\begin{equation}
\nabla\cdot\mathbf B=\mathbf k\cdot\mathbf E,\label{12}
\end{equation}
\begin{equation}
\nabla\cdot\mathbf E=-\mathbf k\cdot\mathbf B.\label{13}
\end{equation}
The set of equations (\ref{5}), (\ref{6}),(\ref{12}) and (\ref{13}) can be
considered as the basic set of  equations of  modified electrodynamics. As
distinct from MCS-electrodynamics the  equations (\ref{6}) and (\ref{12}) 
contain the additional terms describing the  induced magnetic charges and
currents.
\section{The fundamental equations of extended symmetrical classical
electrodynamics}
Let us take the divergence of both sides of (\ref{5}) and (\ref{6}):
\begin{equation}
\frac{1}{c}\frac{\partial}{\partial t}(\nabla\cdot\mathbf E-\mathbf
k\cdot\mathbf B)-m(\nabla\cdot\mathbf B+\mathbf k\cdot\mathbf E)=0,\label{20}
\end{equation}
\begin{equation}
\frac{1}{c}\frac{\partial}{\partial t}(\nabla\cdot\mathbf B+\mathbf
k\cdot\mathbf E)+m(\nabla\cdot\mathbf E-\mathbf k\cdot\mathbf B)=0.\label{21}
\end{equation}
The equations (\ref{12}) and (\ref{13}) can be written in the form:
\begin{equation}
\nabla\cdot\mathbf B=4\pi\rho_m,\label{16}
\end{equation}
\begin{equation}
\nabla\cdot\mathbf E=4\pi\rho_e,\label{17}
\end{equation}
where $\rho_m$ is the density of the induced magnetic charge:
\begin{equation}
\rho_m=\frac{1}{4\pi}\mathbf k\cdot\mathbf E,\label{18}
\end{equation}
and $\rho_e$ is the density of the induced electric charge:
\begin{equation}
\rho_e=-\frac{1}{4\pi}\mathbf k\cdot\mathbf B.\label{19}
\end{equation}
 Substituting (\ref{16})-(\ref{19}) into (\ref{20}) and (\ref{21}), we obtain:
\begin{equation}
\frac{\partial\rho_e}{\partial t}-mc\rho_m=0,\label{22}
\end{equation}
\begin{equation}
\frac{\partial\rho_m}{\partial t}+mc\rho_e=0.\label{23}
\end{equation} 
From (\ref{22}) and (\ref{23}) it follows that the oscillations of the densities
of the induced charges  must occur if $m\ne0$.  The frequency $\omega$ of such
oscillations can be written as $\omega=|m|c$. The oscillations of $\rho_e$
 and $\rho_m$  have the same magnitude but differ in phase by $90^\circ$.
Consequently, an electric charge transforms into a magnetic charge and vice
versa. In this case the law of charge conservation is broken and  therefore we
let $m=0$. With $m=0$ the equations (\ref{22}) and (\ref{23}) become:
\begin{equation}
\frac{\partial\rho_e}{\partial t}=-\frac{1}{4\pi}\mathbf
k\cdot\frac{\partial\mathbf B}{\partial t}=0,\label{24}
\end{equation} 
\begin{equation}
\frac{\partial\rho_m}{\partial t}=\frac{1}{4\pi}\mathbf
k\cdot\frac{\partial\mathbf E}{\partial t}=0.\label{25}
\end{equation}
The conditions (\ref{24}) and (\ref{25}) mean that either the electric and
magnetic fields do not  depend on time or the time-varying $\mathbf B$ and
$\mathbf E$ are both perpendicular to the direction of  $\mathbf k$.

If $m=0$ and  $\mathbf k\ne0$, the set of the equations (\ref{5}), (\ref{6}),
(\ref{12}) and (\ref{13}) can be written
\begin{equation}
\nabla\times\mathbf B-\frac{1}{c}\frac{\partial\mathbf E}{\partial t}=-\mathbf
k\times\mathbf E,\label{26}
\end{equation}
\begin{equation}
\nabla\cdot\mathbf E=-\mathbf k\cdot\mathbf B,\label{27}
\end{equation}
\begin{equation}
\nabla\times\mathbf E+\frac{1}{c}\frac{\partial\mathbf B}{\partial t}=\mathbf
k\times\mathbf B,\label{28}
\end{equation}
\begin{equation}
\nabla\cdot\mathbf B=\mathbf k\cdot\mathbf E.\label{29}
\end{equation}
We call the equations (\ref{26})-(\ref{29})  the fundamental equations of
extended symmetrical classical electrodynamics (ESC- electrodynamics). The
equations (\ref{26})-(\ref{29}) were postulated by  authors in the article 
\cite{fedorov1}.
Computing the curls of the equations (\ref{26}) and (\ref{28}) and using
(\ref{27}) and (\ref{29}), we find:
\begin{equation}
\square\mathbf B+\frac{2}{c}\mathbf k\times\frac{\partial\mathbf B}{\partial
t}+k^2\mathbf B=0,\label{DB} 
\end{equation}
\begin{equation}
\square\mathbf E+\frac{2}{c}\mathbf k\times\frac{\partial\mathbf E}{\partial
t}+k^2\mathbf E=0.\label{DE} 
\end{equation}

The set of the equations (\ref{26})-(\ref{29}) or (\ref{DB})-(\ref{DE})
describes the behavior of the electromagnetic field  in the  space region  where
the vector $\mathbf k$ is not zero.   In general,  the induced charge and
current are distributed over  this region and we call, therefore,  this region 
the induced charge and current domain (ICC-domain). Outside  the ICC-domain, the
behavior of the electromagnetic field is governed by the Maxwell equations. 

\section{The potentials in ESC-electrodynamics}

In ESC-electrodynamics we can define $\mathbf E$ and $\mathbf B$ in terms of
potentials: 
\begin{equation}
\mathbf E=-\nabla\phi'-\nabla\times\mathbf A''-\frac{1}{c}\frac{\partial\mathbf
A'}{\partial t}+\mathbf k\phi''+\mathbf k\times\mathbf A',\label{30}
\end{equation}
\begin{equation}
\mathbf B=-\nabla\phi''+\nabla\times\mathbf A'-\frac{1}{c}\frac{\partial\mathbf
A''}{\partial t}-\mathbf k\phi'+\mathbf k\times\mathbf A'',\label{31}
\end{equation}
where $\phi'$ and $\phi''$ are scalar potentials, $\mathbf A'$ and $\mathbf A''$
are vector potentials. Lorentz covariance requires that the potentials $\phi'$,
$\mathbf A'$, $\phi''$ and $\mathbf A''$ form two 4-vector potentials
\begin{equation}
A'^\mu=(\phi',\mathbf A'), \label{p'}
\end{equation}
\begin{equation}
A''^\mu=(\phi'',\mathbf A''), \label{p''}
\end{equation}
where we use Greek indices $\mu,\nu,...$ to run from 0 to 3. Since $\mathbf E$
is a polar vector and $\mathbf B$ is an axial vector than the components of
$\mathbf E$ and $\mathbf B$ can be  interpreted as the components of the
antisymmetric second rank  tensor\cite{landau2}:
\begin{equation}
F_{\mu\nu}=\left(\begin{array}{cccc}
0&E_x&E_y&E_z\\&0&-B_z&B_y\\&&0&-B_x\\&&&0\end{array}\right).\label{54}
\end{equation} 
We can write the components of $F_{\mu\nu}$ as
\begin{equation}
F_{\mu\nu}=F'_{\mu\nu}-\frac{1}{2}\epsilon_{\mu\nu\rho\sigma}F''^{\rho\sigma},
\label{fmn}
\end{equation}
where
\begin{equation}
F'_{\mu\nu}=\partial_\mu A'_\nu-\partial_\nu A'_\mu+K_{\mu\nu\alpha}A'^\alpha,
\label{f'mn} 
\end{equation}
\begin{equation}
F''^{\rho\sigma}=\partial^\rho A''^\sigma-\partial^\sigma
A''^\rho+K^{\rho\sigma\alpha}A''_\alpha\label{f''mn} 
\end{equation}
and $\epsilon_{\mu\nu\rho\sigma}$ is the totally antisymmetric fourth rank
tensor \cite{landau2}. The tensor $K_{\mu\nu\alpha}$ is a 4-tensor of the third
rank and its nonvanishing components in the rest frame of the ICC-domain are
\begin{equation}
k_{032}=k_{203}=k_{230}=-k_{302}=-k_{023}=-k_{320}=k_x,\label{kx}
\end{equation}
\begin{equation}
k_{013}=k_{301}=k_{310}=-k_{103}=-k_{031}=-k_{130}=k_y,\label{ky}
\end{equation}
\begin{equation}
k_{021}=k_{102}=k_{120}=-k_{201}=-k_{012}=-k_{210}=k_z,\label{kz}
\end{equation}
where the quantities $k_x$, $k_y$ and $k_z$ are the components of the vector
$\mathbf k$. The tensor
$K_{\mu\nu\alpha}$ is  antisymmetric in indices $\mu$ and $\nu$. It can be shown
that
\begin{equation}
K_{\mu\nu\alpha}K^{\mu\nu\alpha}=6(k_x^2+k_y^2+k_z^2)=6k^2.\label{k2} 
\end{equation}
Thus $k^2$ is a Lorentz invariant. With $\mathbf k=0$ the tensor (\ref{fmn}) is
the  Cabibbo-Ferrari tensor \cite{cabibbo}:
\begin{equation}
F_{\mu\nu}=\partial_\mu A'_\nu-\partial_\nu
A'_\mu-\epsilon_{\mu\nu\rho\sigma}\partial^\rho A''^\sigma.\label{cab} 
\end{equation}
Let us write the equations (\ref{26}) and (\ref{28}) as
\begin{equation}
\nabla\times\mathbf B-\frac{1}{c}\frac{\partial\mathbf E}{\partial
t}=\frac{4\pi}{c}\mathbf j_e,\label{95}
\end{equation}
\begin{equation}
\nabla\times\mathbf E+\frac{1}{c}\frac{\partial\mathbf B}{\partial
t}=-\frac{4\pi}{c}\mathbf j_m,\label{97}
\end{equation}
where $\mathbf j_e$ is the density of the induced electric current
\begin{equation}
\mathbf j_e=-\frac{c}{4\pi}\mathbf k\times\mathbf E,\label{99}
\end{equation}
and $\mathbf j_m$ is the  density of the induced magnetic current
\begin{equation}
\mathbf j_m=-\frac{c}{4\pi}\mathbf k\times\mathbf B.\label{100}
\end{equation}
Then we introduce a 4-vector $j_e^\alpha$ as  
\begin{equation}
j_e^\alpha=\frac{c}{8\pi}K^{\mu\nu\alpha}F_{\mu\nu}.\label{je} 
\end{equation}
If the expressions (\ref{54}) and (\ref{kx})-(\ref{kz}) are substituted  into
(\ref{je}), the following expression can be obtained
\begin{equation}
j_e^\alpha=(c\rho_e,\mathbf j_e),\label{je1} 
\end{equation}
where $\rho_e$ and $\mathbf j_e$ are defined by the expressions (\ref{19}) and
(\ref{99}), respectively. Thus the induced electric charge and current densities
are the components of a 4-vector.  We now introduce a 4-vector $j_m^\alpha$ as 
\begin{equation}
j_m^\alpha=\frac{c}{8\pi}K^{\mu\nu\alpha}F^*_{\mu \nu}, \label{jm} 
\end{equation}
where $F^*_{\mu \nu}$ is the dual electromagnetic tensor
\begin{equation}
F^*_{\mu \nu}=\frac{1}{2}\epsilon_{\mu\nu\rho\sigma}F^{\rho\sigma}.\label{F*mn} 
\end{equation}
The expression (\ref{jm}) can be written as
\begin{equation}
j_m^\alpha=(c\rho_m,\mathbf j_m),\label{jm1} 
\end{equation}
where $\rho_m$ and $\mathbf j_m$ are defined by the expressions (\ref{19}) and
(\ref{100}), respectively. Thus the induced magnetic charge and current
densities are also the components of a 4-vector.

The Lagrangian of ESC-electrodynamics can be taken in the form 
\begin{equation}
\mathcal
L=-\frac{1}{16\pi}F_{\mu\nu}F^{\mu\nu}=\frac{1}{8\pi}(E^2-B^2).\label{60} 
\end{equation}
Substituting (\ref{30}) and (\ref{31}) into (\ref{60}) and using the  principle
of least action \cite{landau2}, we  obtain the basic equations
(\ref{26})-(\ref{29}) of ESC-electrodynamics.

\section{The equations for the potentials}

 Substituting (\ref{30}) and (\ref{31})   into (\ref{26}) and (\ref{28}), we
obtain :
\begin{equation}
\square\mathbf A'+\frac{2}{c}\mathbf k\times\frac{\partial\mathbf A'}{\partial
t}+k^2\mathbf A'=\nabla\psi_1+\mathbf k\psi_2,\label{32}
\end{equation}
\begin{equation}
\square\mathbf A''+\frac{2}{c}\mathbf k\times\frac{\partial\mathbf A''}{\partial
t}+k^2\mathbf A''=\nabla\psi_3-\mathbf k\psi_4,\label{33}
\end{equation}
\begin{equation}
\nabla^2\phi'+\frac{1}{c}\frac{\partial}{\partial t}(\nabla \cdot\mathbf
A'+\mathbf k\cdot\mathbf A'')+k^2\phi'=0,\label{phi'} 
\end{equation}
\begin{equation}
\nabla^2\phi''+\frac{1}{c}\frac{\partial}{\partial t}(\nabla \cdot\mathbf
A''-\mathbf k\cdot\mathbf A')+k^2\phi''=0,\label{phi''} 
\end{equation}
where
\begin{equation}
\psi_1=\nabla\cdot\mathbf A'+\frac{1}{c}\frac{\partial\phi'}{\partial t}-\mathbf
k\cdot\mathbf A'',\label{34} 
\end{equation}
\begin{equation}
\psi_2=\nabla\cdot\mathbf A''-\frac{1}{c}\frac{\partial\phi''}{\partial
t}+\mathbf k\cdot\mathbf A',\label{35} 
\end{equation}
\begin{equation}
\psi_3=\nabla\cdot\mathbf A''+\frac{1}{c}\frac{\partial\phi''}{\partial
t}+\mathbf k\cdot\mathbf A',\label{36} 
\end{equation}
\begin{equation}
\psi_4=\nabla\cdot\mathbf A'-\frac{1}{c}\frac{\partial\phi'}{\partial t}-\mathbf
k\cdot\mathbf A''.\label{37} 
\end{equation}
In order to determine the divergence of  $\mathbf A'$ and $\mathbf A''$, we can
use the expressions (\ref{34}) and (\ref{36}).  We assume that $\psi_1=0$ and
$\psi_3=0$. In this case we have 
\begin{equation}
\nabla\cdot\mathbf A'+\frac{1}{c}\frac{\partial\phi'}{\partial t}=\mathbf
k\cdot\mathbf A'',\label{38} 
\end{equation}
\begin{equation}
\nabla\cdot\mathbf A''+\frac{1}{c}\frac{\partial\phi''}{\partial t}=-\mathbf
k\cdot\mathbf A'.\label{39} 
\end{equation}
Substituting (\ref{38}) and (\ref{39}) into (\ref{32})-(\ref{phi''}), we
obtain: 
\begin{equation}
\square\mathbf A'+\frac{2}{c}\mathbf k\times\frac{\partial\mathbf A'}{\partial
t}+k^2\mathbf A'=-\frac{2}{c}\mathbf k\frac{\partial\phi''}{\partial
t},\label{42}
\end{equation}
\begin{equation}
\square\mathbf A''+\frac{2}{c}\mathbf k\times\frac{\partial\mathbf A''}{\partial
t}+k^2\mathbf A''=\frac{2}{c}\mathbf k\frac{\partial\phi'}{\partial
t}.\label{43}
\end{equation}
\begin{equation}
\square\phi'+k^2\phi'=-\frac{2}{c}\mathbf k\cdot\frac{\partial\mathbf
A''}{\partial t},\label{fi'}
\end{equation}
\begin{equation}
\square\phi''+k^2\phi''=\frac{2}{c}\mathbf k\cdot\frac{\partial\mathbf
A'}{\partial t}.\label{fi''}
\end{equation}
The set of the equations (\ref{38})- (\ref{fi''}) is equivalent to the set of
the equations (\ref{26})-(\ref{29}).

Let us examine the case in which the electromagnetic field  is  defined in terms
of only one 4-potential. Let $A'^\mu \ne0$ and $A''^\mu=0$. From the equations
(\ref{38})-(\ref{fi''})  it follows that the following conditions must be
satisfied
\begin{equation}
\nabla\cdot\mathbf A'+\frac{1}{c}\frac{\partial\phi'}{\partial t}=0,\label{LA'} 
\end{equation} 
\begin{equation}
\mathbf k\cdot\mathbf A'=0,\label{kA'} 
\end{equation}
\begin{equation}
\frac{\partial\phi'}{\partial t}=0.\label{df't} 
\end{equation}
Substituting (\ref{df't}) into (\ref{LA'}), we find
\begin{equation}
\nabla\cdot\mathbf A'=0.\label{CA'} 
\end{equation}
Thus the vector-potential $\mathbf A'$  satisfies   the Coulomb gauge condition.
 The scalar potential $\phi'$ of the time-varying electromagnetic field
vanishes. Consequently, the time-varying electromagnetic field is governed by
the equation
\begin{equation}
\square\mathbf A'+\frac{2}{c}\mathbf k\times\frac{\partial\mathbf A'}{\partial
t}+k^2\mathbf A'=0.\label{DA'}
\end{equation}
From (\ref{30}) and (\ref{31}) it follows that the vectors of the time-varying
electromagnetic field are 
\begin{equation}
\mathbf E=-\frac{1}{c}\frac{\partial\mathbf  A'}{\partial t}+\mathbf
k\times\mathbf A',\label{Et} 
\end{equation}
\begin{equation}
\mathbf B=\nabla\times \mathbf A'.\label{Bt} 
\end{equation}

In the static case the electromagnetic field is  determined  by the scalar
potential $\phi'$ and the vector-potential $\mathbf A'$, which are the solutions
of the equations:
\begin{equation}
\nabla^2\mathbf A'+k^2\mathbf A'=0,\label{A's} 
\end{equation}
\begin{equation}
\nabla^2\phi'+k^2\phi'=0,\label{f's} 
\end{equation}
  The vectors of the static field can be  obtained  from (\ref{30}) and
(\ref{31}), which are in this case:
\begin{equation}
\mathbf E=-\nabla\phi'+\mathbf k\times\mathbf A',\label{Ec} 
\end{equation}
\begin{equation}
\mathbf B=\nabla\times \mathbf A'-\mathbf k\phi'.\label{Bc} 
\end{equation}

The equation similar to (\ref{f's}) is used to calculate the ``self-induced'' 
electrostatic  field in the paper \cite{kapus}. We can  find similar expressions
for the 4-potential $A''^\mu$ if $A''^\mu\ne0$ and $A'^\mu=0$. 
\section{The static electromagnetic field in ESC-electrodynamics and the
electric monopole}
In the static case the relations (\ref{24}) and (\ref{25}) are satisfied
automatically and the set of the basic equations of ESC-electrodynamics becomes
\begin{equation}
\nabla\times\mathbf B=-\mathbf k\times\mathbf E,\label{61}
\end{equation}
\begin{equation}
\nabla\cdot\mathbf E=-\mathbf k\cdot\mathbf B,\label{62}
\end{equation}
\begin{equation}
\nabla\times\mathbf E=\mathbf k\times\mathbf B,\label{63}
\end{equation}
\begin{equation}
\nabla\cdot\mathbf B=\mathbf k\cdot\mathbf E.\label{64}
\end{equation}
Combining the curls of the equations (\ref{61}) and (\ref{63}), we obtain the
equations:
\begin{equation}
\Delta\mathbf B+k^2\mathbf B=0\label{65}, 
\end{equation}
\begin{equation}
\Delta\mathbf E+k^2\mathbf E=0\label{66}. 
\end{equation}

If we find the magnetic field from (\ref{65}) then the electric field can be
calculated from (\ref{61})-(\ref{64}):
\begin{equation}
\mathbf E=\frac{1}{k^2}\{\mathbf k\times(\nabla\times\mathbf B)+\mathbf
k(\nabla\cdot\mathbf B)\}\label{67}.
\end{equation}
We call such a state of the electromagnetic field   the state of the magnetic
type.  We can similarly find the electric field from (\ref{66}) and then the
magnetic field can be calculated from (\ref{61})-(\ref{64}):
\begin{equation}
\mathbf B=-\frac{1}{k^2}\{\mathbf k\times(\nabla\times\mathbf E)+\mathbf
k(\nabla\cdot\mathbf E)\}\label{68}.
\end{equation}
We call the  electromagnetic field  described by  (\ref{66}) and (\ref{68})  the
state of the electric type.
\subsection{The electromagnetic field of the electric monopole}
Let us now show that the  set (\ref{61})-(\ref{64}) have the solution which
describes the electromagnetic field of the electric monopole. Such a state of
the electromagnetic field is the state of the electric type and it is defined by
the equations (\ref{66}) and (\ref{68}). Then we assume that the ICC-domain is
the sphere of radius $R$ which is at rest with respect to an  inertial frame of
reference.  In this case it is convenient to use spherical coordinates
$(r,\theta,\alpha)$  where $\theta$ is the angle between $\mathbf k$ and
$\mathbf r$. We  denote the unit vectors of the spherical coordinate system by 
$\mathbf e_r$, $\mathbf e_\theta$ and $\mathbf e_\alpha$. 

The electric field of the electric monopole has spherical symmetry. In the
ICC-domain of the monopole $(r< R)$ such the solution of the equation (\ref{66})
can be written as
\begin{equation}
\mathbf E=\frac{C}{\sqrt{kr}}J_{\frac{3}{2}}(kr)\mathbf e_r,\label{69}
\end{equation}
where $C$ is a constant, $J_{\frac{3}{2}}(kr)$ is the Bessel function.  The
static Maxwell equations are used to find the electromagnetic field outside the
ICC-domain. In the outer region  of the monopole $(r>R)$ the electric field
$\mathbf E^e$ is the solution of the Laplace equation:
\begin{equation}
\mathbf E^e=\frac{q}{r^2}\mathbf e_r\label{70}, 
\end{equation}
where $q$ is the electric charge of the monopole. Substituting (\ref{69}) in
(\ref{68}) we find that the  magnetic field in the inner region $(r< R)$ of the
monopole is given by
\begin{equation}
\mathbf B=-\frac{C}{\sqrt{kr}}J_{\frac{1}{2}}(kr)\mathbf e_k,\label{71}
\end{equation}
where $\mathbf e_k$ is an unit vector parallel to $\mathbf k$. The magnetic
field components are:
\begin{equation}
B_r=-\frac{C}{\sqrt{kr}}J_{\frac{1}{2}}(kr)\cos\theta,\label{72}
\end{equation}
\begin{equation}
B_\theta=\frac{C}{\sqrt{kr}}J_{\frac{1}{2}}(kr)\sin\theta.\label{73}
\end{equation}
From (\ref{71})-(\ref{73}) it is seen that  the magnetic field in the ICC-domain
is a dipole field.  In the outer region $(r>R)$ the magnetic dipole field is a
solution of the Laplace equation and its components are
\begin{equation}
B_r^e=\frac{2\mu}{r^3}\cos\theta,\label{74}
\end{equation} 
\begin{equation}
B_\theta^e=\frac{\mu}{r^3}\sin\theta,\label{75}
\end{equation}
where $\mu$ is the magnetic moment of the monopole. Let us assume that the
magnetic field must be continuous at the interface between the inner and outer
regions of the monopole. In our case this condition is satisfied if the magnetic
field vanishes  at $r\ge R$. From (\ref{72})-(\ref{73}) it follows that the
boundary condition for the magnetic field can be satisfied   only if
\begin{equation}
\sin(kR)=0.\label{76}
\end{equation}
This means that $k$ have the eigenvalues:
\begin{equation}
k_n^+=n\frac{\pi}{R},\label{77}
\end{equation}
\begin{equation}
k_n^-=-n\frac{\pi}{R},\label{k_n}
\end{equation}
\begin{equation}
n=0,1,2,3,...\label{n}
\end{equation}
Consequently, the eigenvalues of $\mathbf k$ are
\begin{equation}
\mathbf k_n^+=n\frac{\pi}{R}\mathbf e_k,\label{kn+}
\end{equation}
\begin{equation}
\mathbf k_n^-=-n\frac{\pi}{R}\mathbf e_k.\label{kn-}
\end{equation}
Thus the two eigenvalues $k_n^+$ and $k_n^-$ correspond to the given  value of
$n$. The solution  corresponding $k_n^+$ we call  the positive component of the
$n$-th mode and the solution corresponding $k_n^-$ we call  the negative
component of the $n$-th mode. The electric and magnetic vectors of the positive
or negative components of $n$-th mode can be written as 
\begin{equation}
\mathbf E_n^\pm=\pm\frac{C_n^\pm}{\sqrt{k_n^+r}}J_{\frac{3}{2}}(k_n^+r)\mathbf
e_r,\label{79}
\end{equation}
\begin{equation}
\mathbf B_n^\pm=-\frac{C_n^\pm}{\sqrt{k_n^+r}}J_{\frac{1}{2}}(k_n^+r)\mathbf
e_k.\label{80}
\end{equation}
In accordance with the superposition principle the electric and magnetic fields
of the $n$-th mode are given by
\begin{equation}
\mathbf E_n=\mathbf E_n^++\mathbf
E_n^-=\frac{Q_n}{R^2\sqrt{k_n^+r}}J_{\frac{3}{2}}(k_n^+r)\mathbf e_r,\label{81}
\end{equation}
\begin{equation}
\mathbf B_n=\mathbf B_n^++\mathbf
B_n^-=-\frac{G_n}{R^2\sqrt{k_n^+r}}J_{\frac{1}{2}}(k_n^+r)\mathbf e_k,\label{82}
\end{equation}
where constants
\begin{equation}
Q_n=(C_n^+-C_n^-)R^2,\label{83}
\end{equation} 
\begin{equation}
G_n=(C_n^++C_n^-)R^2\label{84}
\end{equation}
have dimensions of  electric or magnetic charge correspondingly. In the outer
region of the monopole $(r>R)$ the electric field of the $n$-th mode is 
\begin{equation}
\mathbf E_n^e=\frac{q_n}{r^2}\mathbf e_r\label{85},
\end{equation} 
where  $q_n$ is the electric charge of the $n$-th mode of the monopole. With
$r=R$ the electric field is continuous and, therefore, from (\ref{81}) and
(\ref{85}) it  follows
\begin{equation}
Q_n=-\pi\sqrt{\frac{\pi}{2}}n(-1)^nq_n.\label{86}
\end{equation}
Thus the electric field of the monopole have only  the radial  component. The
dipole magnetic field is non-zero  in the ICC-domain of the monopole.

 In accordance with the superposition principle the vector $\mathbf k$ is an
operator that acts on the vectors $\mathbf E$  and $\mathbf B$ as follows
\begin{equation}
\mathbf k\cdot\mathbf F=\sum_n(\mathbf k_n^+\cdot\mathbf F_n^++\mathbf
k_n^-\cdot\mathbf F_n^-)=\sum_n\mathbf k_n^+\cdot(\mathbf F_n^+-\mathbf
F_n^-)=\sum_n\mathbf k_n^+\cdot\tilde {\mathbf F}_n,\label{87}
\end{equation}
\begin{equation}
\mathbf k\times\mathbf F=\sum_n(\mathbf k_n^+\times\mathbf F_n^++\mathbf
k_n^-\times\mathbf F_n^-)=\sum_n\mathbf k_n^+\times(\mathbf F_n^+-\mathbf
F_n^-)=\sum_n\mathbf k_n^+\times\tilde {\mathbf F}_n,\label{88}
\end{equation}
where $\mathbf F$ is the vector $\mathbf E$ or $\mathbf B$ and 
\begin{equation}
\tilde {\mathbf F}_n=\mathbf F_n^+-\mathbf F_n^-\label{90}.
\end{equation}
Substituting the expressions (\ref{79}) and (\ref{80}) in (\ref{90}), we obtain
for the electric monopole
\begin{equation}
\tilde {\mathbf E}_n=\mathbf E_n^+-\mathbf
E_n^-=\frac{G_n}{R^2\sqrt{k_n^+r}}J_{\frac{3}{2}}(k_n^+r)\mathbf e_r,\label{91}
\end{equation}
\begin{equation}
\tilde {\mathbf B}_n=\mathbf B_n^+-\mathbf
B_n^-=-\frac{Q_n}{R^2\sqrt{k_n^+r}}J_{\frac{1}{2}}(k_n^+r)\mathbf e_r.\label{92}
\end{equation}
With $\mathbf k=0$ the equations of ESC-electrodynamics become  the Maxwell
equations. Hence the operator $\mathbf k$ does not act on the vectors of Maxwell
electrodynamics
\begin{equation}
\mathbf k\cdot\mathbf F_{M}=0,\label{93}
\end{equation}
\begin{equation}
\mathbf k\times\mathbf F_{M}=0,\label{94}
\end{equation}
where $\mathbf F_{M}$ is the vector of the Maxwell field. Thus the densities of 
induced charges and currents of the electric monopole do not depend on the
Maxwell field.
\subsection{The energy and the angular moment of the electric monopole}
It can be shown that, similarly to  Maxwell electrodynamics, the   Poynting
theorem of ESC-electrodynamics can be written as follows
\begin{equation}
\frac{1}{8\pi}\frac{\partial w}{\partial t}+\nabla\cdot\mathbf S=-\mathbf
E\cdot\mathbf j_e-\mathbf B\cdot\mathbf j_m,\label{101}
\end{equation}
where $w$ is the electromagnetic energy density:
\begin{equation}
w=\frac{1}{8\pi}(\mathbf E^2+\mathbf B^2)\label{102}
\end{equation}
and $\mathbf S$ is the Poynting vector: 
\begin{equation}
\mathbf S=\frac{c}{4\pi}\mathbf E\times\mathbf B.\label{103}
\end{equation}
Since the electromagnetic field of the monopole is static, the Poynting theorem
can be written as 
\begin{equation}
\nabla\cdot\mathbf S=-\mathbf E\cdot\mathbf j_e-\mathbf B\cdot\mathbf
j_m.\label{104}
\end{equation}
Substituting (\ref{81}) and (\ref{82}) in (\ref{103}), we obtain 
\begin{equation}
\mathbf S=\sum_{n,p}\frac{cQ_nG_p}{4\pi
R^4r\sqrt{k_n^+k_p^+}}J_{\frac{3}{2}}(k_n^+r)J_{\frac{1}{2}}(k_p^+r)\sin
\theta\mathbf e_\alpha.\label{105}
\end{equation}
Since the value of the Poynting vector is independent on the angle $\alpha$ the
left-hand side of (\ref{104}) vanishes. It is seen from (\ref{99}) that the
vector $\mathbf j_e$  is perpendicular to the vector $\mathbf E$, hence the
first term on the right-hand side of (\ref{104}) is zero. Vector $\mathbf B$ is
collinear to $\mathbf k$, hence $\mathbf j_m=0$ and the second term on the
right-hand side of (\ref{104}) vanishes. 
 We see that the Poynting theorem is valid in the inner region of the monopole.
In the outer region of the monopole Poynting's theorem is also  valid as
$\mathbf S$, $\mathbf j_e$ and $\mathbf j_m$ are zero in this case.

From (\ref{102}) we have that the electric energy density can be written in the
form
\begin{equation}
w_e=\frac{\mathbf E^2}{8\pi}\label{108}
\end{equation}
and the electric energy of the monopole is
\begin{equation}
W_e=\frac{1}{8\pi}\sum_{n,p}\int_V\mathbf E_n\cdot\mathbf E_pdV \label{We} 
\end{equation}
Substituting (\ref{81}) in (\ref{We}) and integrating over the volume $V^{in}$
of the ICC-domain, we find that electric energy of the inner region of the
monopole is
\begin{equation}
W_e^{in}=\sum_{n}(W_e^{in})_n,\label{110}
\end{equation}
where $(W_e^{in})_n$ is the electric energy of the $n$-th mode of the inner
region of the monopole:
\begin{equation}
(W_e^{in})_n=\frac{Q_n^2}{2\pi^3n^2R}=\frac{q_n^2}{4R}.\label{111}
\end{equation}
Substituting (\ref{85}) in (\ref{We}) and integrating over the volume $V^{e}$ of
the outer region of the monopole, we find that electric energy of the external
region of the monopole is
\begin{equation}
W_e^e=\sum_{n,p}\frac{q_nq_p}{2R}=\sum_{n,p}\frac{(-1)^{n+p}Q_nQ_p}{\pi^3npR}
.\label{112}
\end{equation}
Thus the electric energy of the $n$-th mode of the outer region of the monopole
is
\begin{equation}
(W_e^e)_n=\frac{q_n^2}{2R}=\frac{Q_n^2}{\pi^3n^2R}.\label{113}
\end{equation} 
Summing (\ref{111}) and (\ref{113}) we find that the total electric energy of
$n$-th mode of the monopole is
\begin{equation}
(W_e)_n=\frac{3Q_n^2}{2\pi^3n^2R}=\frac{3q_n^2}{4R}.\label{114}
\end{equation}
The expressions (\ref{112}) also contain  the interaction energy between $n$-th
and $p$-th modes which can be written in the form
\begin{equation}
(W_e^e)_{np}=\frac{q_nq_p}{R}=\frac{2(-1)^{n+p}Q_nQ_p}{\pi^3npR}.\label{115}
\end{equation} 

From (\ref{102}) we have that the magnetic energy density is
\begin{equation}
w_m=\frac{\mathbf B^2}{8\pi}.\label{116}
\end{equation}
and the magnetic energy of the monopole is
\begin{equation}
W_m=\frac{1}{8\pi}\sum_{n,p}\int_V\mathbf B_n\cdot\mathbf B_pdV \label{Wm} 
\end{equation}
Substituting (\ref{82}) in (\ref{Wm}) and integrating over the volume of the
ICC-domain, we find that magnetic energy of the inner region of the monopole is
\begin{equation}
W_m^{in}=\sum_{n}(W_m^{in})_n,\label{118}
\end{equation}
where $(W_m^{in})_n$ is the magnetic energy of the $n$-th mode of the inner
region of the monopole
\begin{equation}
(W_m^{in})_n=\frac{G_n^2}{2\pi^3n^2R}.\label{119}
\end{equation}
The magnetic field and the magnetic energy are zero in the outer region of the
monopole.

Summing (\ref{114}) and (\ref{119}) we find that the total electromagnetic
energy of $n$-th mode of the monopole is
\begin{equation}
W_n=\frac{3Q_n^2+G_n^2}{2\pi^3n^2R}=\frac{3q_n^2}{4R}+\frac{G_n^2}{2\pi^3n^2R}
.\label{120}
\end{equation}

In ESC-electrodynamics,  the electromagnetic momentum density can be written in
the standard form 
\begin{equation}
\mathbf p=\frac{1}{4\pi c}\mathbf E\times \mathbf B .\label{121} 
\end{equation}
Hence the angular momentum  of the monopole is
\begin{equation}
\mathbf J=\frac{1}{4\pi c}\sum_{n,p}\int_V(\mathbf r\times (\mathbf E_n\times
\mathbf B_p))dV .\label{123} 
\end{equation}
Substituting (\ref{81}) and (\ref{82}) in (\ref{123}) and integrating over the
volume of the ICC-domain , we find that the angular momentum of $n$-th mode is 
\begin{equation}
\mathbf J_n=\frac{Q_nG_n}{\pi^4cn^3}\mathbf e_k .\label{124} 
\end{equation}
There are the components in (\ref{123}), which contain the electric field of the
$n$-th mode and the magnetic field of the $p$-th mode and vice versa. We call
 these components  the cross angular momenta of the $n$-th and $p$-th modes. In
our case ones can be written in the form
\begin{equation}
\mathbf J_{np}=\frac{4(nQ_pG_n-pQ_nG_p)(-1)^{n+p}}{3\pi^4cnp(n^2-p^2)}\mathbf
e_k .\label{125} 
\end{equation}
Now consider a particular case. If the constants $Q_n$ and  $G_n$  satisfy to
the conditions
\begin{equation}
Q_n=-(-1)^nnQ,\label{126}
\end{equation}
\begin{equation}
G_n=-(-1)^nn^2G,\label{127}
\end{equation}
then the angular momenta (\ref{124}) and (\ref{125}) become
\begin{equation}
\mathbf J_n=\frac{QG}{\pi^4c}\mathbf e_k,\label{128} 
\end{equation}
\begin{equation}
\mathbf J_{np}=\frac{4QG}{3\pi^4c}\mathbf e_k.\label{129}
\end{equation}
Substituting (\ref{126}) in (\ref{86}), we have
\begin{equation}
q_n=q=\frac{1}{\pi}\sqrt{\frac{2}{\pi}}Q.\label{130}
\end{equation}

Suppose that there is only one mode of the electromagnetic field  in the
ICC-domain. It follows   from the expressions (\ref{128}) and (\ref{130}) that
the angular momentum and charge of the monopole  do not depend on the state
number $n$.  From (\ref{128}) we see that the projection of the angular momentum
of $n$-th state on the direction of  $\mathbf k$ is 
\begin{equation}
J=\frac{QG}{\pi^4c},\label{131} 
\end{equation}
Then from (\ref{130}) and (\ref{131}) we have that the expression (\ref{120})
can be written in the form
\begin{equation}
W_n=\frac{3q^2}{4R}+\frac{(\pi cnJ)^2}{q^2R}.\label{132}
\end{equation}
If $J$ is constant then the energy minimum  of the $n$-th state is reached   at
a certain magnitude of the electric charge. For example the energy of the first
state of the monopole is  minimum if the electric charge is 
\begin{equation}
q'=\pm \sqrt{\frac{2\pi c|J|}{\sqrt3}}.\label{133}
\end{equation}
In our case   electric charge of any state is equal to the electric charge of
the first state and therefore the energy of any state of the monopole  is
\begin{equation}
W_n=\frac{\sqrt{3}\pi c|J|}{2R}(1+n^2).\label{134}
\end{equation}
With $q=q'$ and $n=1$ the electric energy of the monopole is equal to its
magnetic energy. The electric monopole energy  does not depend on the  state
number $n$. 
Thus we showed that the set of the equations (\ref{61})-(\ref{64}) has the
particular solutions which describe electric monopoles. From  the equations
(\ref{65}) and (\ref{67})  we can obtain the solution which describes magnetic
monopoles. In general, the ICC-domain can have any multipole moment.
\section{Conclusions}
In the present article the governed equations of ESC-electrodynamics  are
formulated in which ``ordinary'' point charges  are absent. The equations of
ESC-electrodynamics contain the additional terms  describing the   induced
charges and currents.    The  densities of the induced charges and currents   
depend on the vector $\mathbf k$ and the vectors of the electromagnetic field
$\mathbf E$ and $\mathbf B$.  

In ESC-electrodynamics the vectors $\mathbf E$ and $\mathbf B$ can be defined in
terms of two 4-potentials. It is shown that the components of $\mathbf k$ are
the components of the 4-tensor of the third rank. The Lagrangian 
   is  found for ESC-electrodynamics. We obtain the conditions for
which there is only one  4-potential describing the behavior of the
electromagnetic field.

It is shown that static ESC-electrodynamics can describe the electromagnetic
field in the inner region of the electric monopole. There are both the electric
and magnetic fields in the inner region. In the outer region of the electric
monopole the electric field is governed by the Maxwell equations. From the
boundary conditions at the interface between the inner and outer regions  it
follows that the vector $\mathbf k$ has a discrete spectrum. The electric and
magnetic fields, energy and angular momentum are found for  different values of
$\mathbf k$. The structure of the electromagnetic field in the inner region of
the monopole   is changed in a discrete way when a transition occurs between
different states.
 
There is a particular case in which different states of the monopole have the
same angular momenta and charges. Such properties of the electric monopole are
similar to properties of elementary particles.

\end{document}